\documentclass{article}
\usepackage{cite}

\usepackage{graphicx}

\title{Isolating electrons on superfluid helium}

\author{Maika Takita and S.A. Lyon\thanks{Department of Electrical Engineering, Princeton University, Princeton, NJ, USA mtakita@princeton.edu}}
\date{}

\begin{document}
\maketitle

\begin{abstract}
Electrons floating on the surface of superfluid helium have been suggested as promising mobile spin quantum bits (qubits).  Transferring electrons extremely efficiently in a narrow channel structure with underlying gates has been demonstrated, showing no transfer error while clocking $10^9$ pixels in a 3-phase charge coupled device (CCD). While on average, one electron per channel was clocked, it is desirable to reliably obtain a single electron per channel. We have designed an electron turnstile consisting of a narrow (0.8$\mu$m) channel and narrow underlying gates (0.5$\mu$m) operating across seventy-eight parallel channels. Initially, we find that more than one electron can be held above the small gates. Underlying gates in the turnstile region allow us to repeatedly split these electron packets. Results show a plateau in the electron signal as a function of the applied gate voltages, indicating quantization of the number of electrons per pixel, simultaneously across the seventy-eight parallel channels.
\end{abstract}

\section{Introduction}
Electrons are bound above the surface of liquid helium by their weak image potential as well as potentials from underlying electrostatic gates \cite{Sommer_1964,Williams_1971,Grimes_1978,Andrei_1997}. These underlying gates control the position of electrons on helium and can move them across the sample into various regions \cite{Bradbury_2011,Sabouret_2008}. The electrons reside in vacuum, about 11nm above the surface, forming a very clean classical two dimensional electron system. Electron spins are expected to have little spin decoherence, due to the small spin-orbit interaction \cite{Lyon_2006}, allowing for the possibility of mobile spin qubits in a condensed matter system. To realize mobile spin qubits, control of individual electrons is necessary. With a single-electron transistor (SET) placed under the helium surface, individual electrons have been counted as they come in and out of a trap \cite{Papageorgiou_2005}. Electron transport of a single line of charge has been observed with a point constriction formed in a microchannel using a split gate \cite{Rees_2011}. We show that in addition to the confinement from a channel, narrow underlying gates perpendicular to the channel can be used to trap electrons over one electrostatic gate. The narrow channel and gates for isolating single electrons and wider channels with CCD's are integrated, moving towards single electron control.

\section{Device Structure}
\begin{figure}[htb]
\begin{center}
\includegraphics[width=\textwidth]{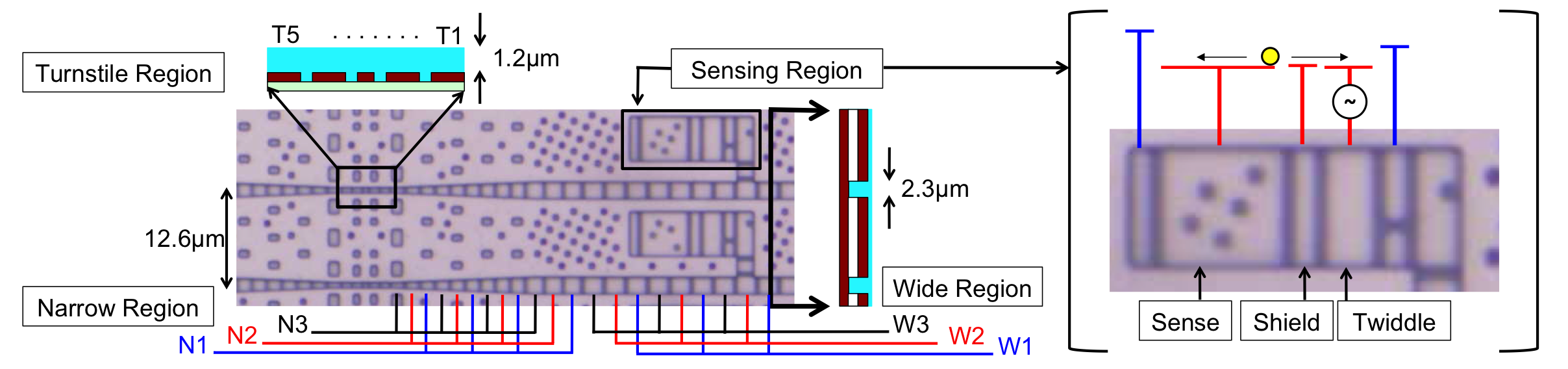}
\caption{\label{fig:pic} Microscope image of the device with cartoon of the channel profile. Two out of seventy-eight channels are shown with gate labels. There are two sets of three-phase CCD's in the narrow region and the wide region. Dark spots on the images are holes etched through the metal to improve the yield of the fabrication processes by having a similar metal density over large areas. An enlarged cartoon of the turnstile cross-section and the side-view of the channels show their dimensions. The cartoon of electron detecting scheme; confining the electrons over sense, shield, and twiddle gate, by keeping neighboring gates repulsive and modulating twiddle gate to move electrons on and off the sense gate; is shown over magnified image of sensing region.}
\end{center}
\end{figure}

The complementary metal-oxide-silicon (CMOS) process is a standard foundry process that consists of multiple metal layers above a silicon substrate, with the metal layers separated by insulators. The helium channel device \cite{Marty_1986,Glasson_2000,Sabouret_2008,Ikegami_2009,Rees_2011} with turnstiles was fabricated utilizing this technology. Seventy-eight parallel channels were designed using one of the metal layers as the top ground plane, with the metal layer below defining the gates that underlay all the channels as shown in Fig.\ref{fig:pic}. The device consists of an electron collecting region to the right of the image in Fig.\ref{fig:pic}, sensing region, wide CCD (WCCD) region, narrow CCD (NCCD) region, and a turnstile region. Electrons are collected over a wide underlying gate, termed \textit{plate}, and clocked into 8.7$\mu$m long pixels (note: three gates comprise a pixel, with a 2.9$\mu$m gate period), both of which have 2.3$\mu$m wide channels. The WCCD region is followed by a NCCD region with four three-phase pixels. The first NCCD has a 7.5$\mu$m pixel length while it remains in a 2.3$\mu$m wide channel. It then necks down to a 0.8$\mu$m wide channel over three 6$\mu$m period pixels. The NCCD section is followed by the turnstile region which contains five individual gates. The first two gates, designated \textit{T1} and \textit{T2}, are 1$\mu$m wide, with 0.4$\mu$m spacing. The center gate, \textit{T3}, is the narrowest gate (0.5$\mu$m), with 0.4$\mu$m spacing to the neighboring gates. The fourth and fifth gates, \textit{T4} and \textit{T5}, have the same dimensions as the first two, which forms a symmetric turnstile region. Electrons are detected in the sensing region which branches out from the channel, creating a larger area for electrons to be detected through a capacitive coupling to a wide \textit{sense} gate \cite{Bradbury_2011}. A small ac-voltage (V$_{ac}$) is applied to the \textit{twiddle} gate to push electrons on and off the \textit{sense}, while a narrow gate between these two gates shields the direct capacitive coupling between them. The periodically induced voltage on the \textit{sense} gate induced by the electrons at the frequency of V$_{ac}$ (111kHz) is buffered by a high electron mobility transistor (HEMT) and detected using a lock-in amplifier. The HEMT is mounted on the backside of the printed circuit board, on which the sample is wirebonded.

\section{Electron Detection}

\begin{figure}[htb]
\begin{center}
\includegraphics[width=\textwidth]{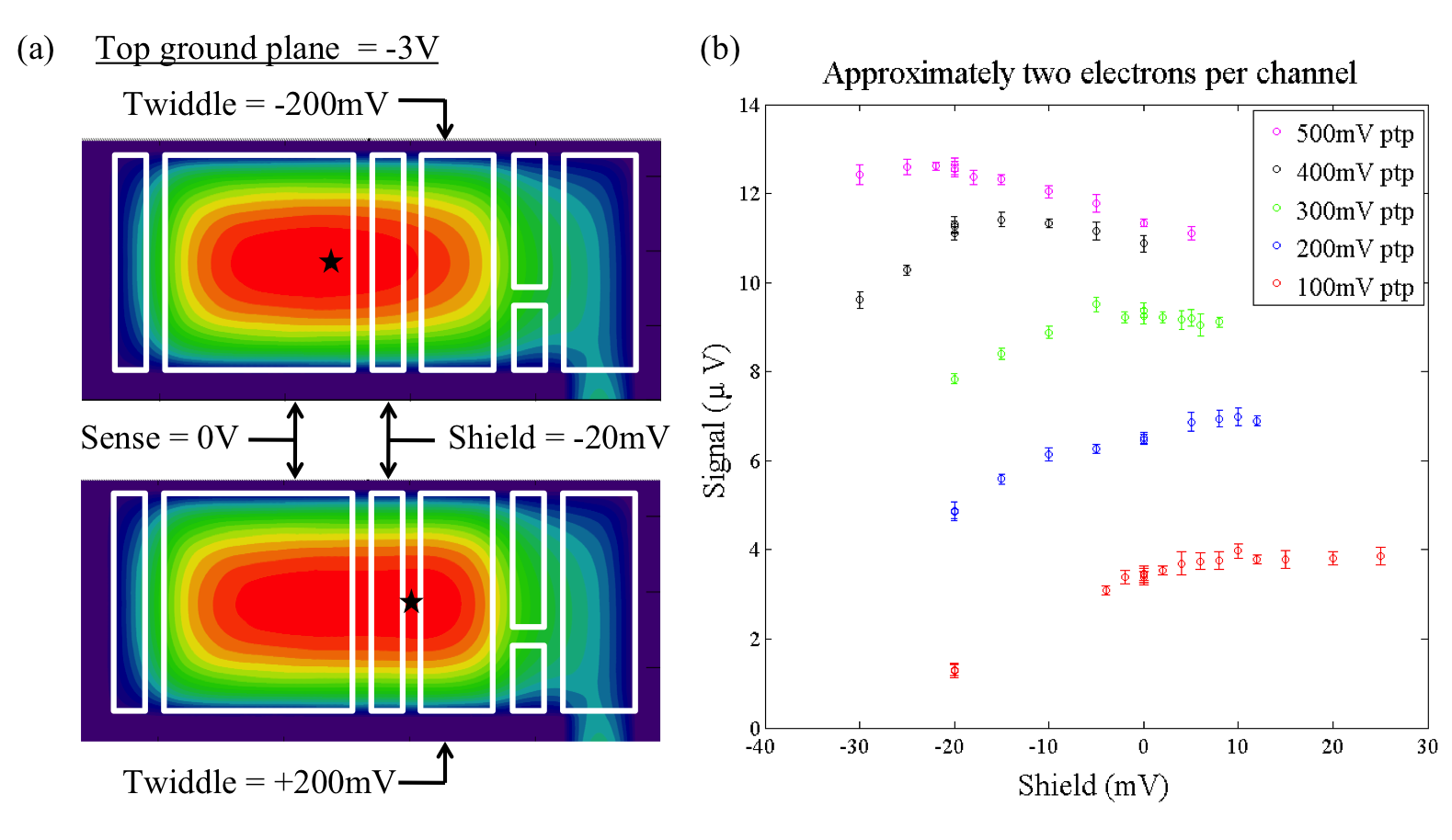}
\caption[MOSIS: FEM for sensing region]{Maximizing electron signal. (a) Potential simulation of the sensing region when electrons are on the \textit{sense} being pushed away from the \textit{twiddle} when it is at the most repulsive potential (-200mV) and when electrons are off the \textit{sense} being pulled over to the \textit{twiddle} when it is at the most attractive potential (+200mV). The gates to the right of the \textit{twiddle} are kept at -2V and the gate to the left of the \textit{sense} is kept at -3V. (b) The plot of electron signal while varying \textit{shield} gate voltage with different V$_{ac}$ on \textit{twiddle}. \label{fig:MOSIS-sense-FEM} }
\end{center}
\end{figure}

The large \textit{sense} gate in the wider channels allows for improved electron detection due to a decreased coupling to the top ground plane. Figure \ref{fig:MOSIS-sense-FEM} (a) shows the potential simulation of the sensing region when electrons are pushed on and off the \textit{sense} gate. The star is placed at the potential energy minimum in each case. When electrons are pushed away from the \textit{twiddle} gate the potential energy minimum is on the \textit{sense} gate. When they are pulled towards the \textit{twiddle}, the potential energy minimum is shifted away from the \textit{sense}. The signal will increase as electrons are brought closer to the center of the \textit{sense} and then pushed completely off the \textit{sense} onto the \textit{twiddle}. This is accomplished by applying a larger V$_{ac}$ on \textit{twiddle}. Figure \ref{fig:MOSIS-sense-FEM} (b) is a plot of measured electron signals from a range of applied voltages on the \textit{shield} gate for different V$_{ac}$ on the \textit{twiddle} gate. The signal increases as the amplitude of V$_{ac}$ increases, confirming that electrons are getting closer to the middle of the \textit{sense} where the induced voltage will be maximized.

The potential energy minimum when electrons are pushed onto the \textit{sense} gate could also be shifted by applying a small negative bias on the \textit{shield} gate placed in between \textit{twiddle} and \textit{sense}. This also shifts the minimum away from the \textit{sense} gate when electrons are pulled onto the \textit{twiddle} gate. The experiments were performed with -20mV on the \textit{shield} gate, and applying a 400mV \textit{twiddle} V$_{ac}$.

\section{Experiment and Results}

Electrons are photoemitted into the vacuum above the \textit{plate} from a zinc film placed a few millimeters above the device \cite{Shankar_2010}. Electrons collected on the \textit{plate} are subsequently clocked into the sensing region using the WCCD gates. Starting with an arbitrary number of electrons per channel, these packets of electrons are brought into the NCCD gates. As the charge packet goes through the NCCD gates, the number of electrons that are allowed on each gate decreases as the channel narrows. Approximately 20 electrons per channel remain above turnstile gate, \textit{T1}, when appropriate voltages are applied. These electrons are our initial charge packet. Other electrons that are not held on \textit{T1} are left behind above other gates, and these are clocked back into the \textit{plate}. The initial packet of electrons are brought onto \textit{T2}, \textit{T3}, and \textit{T4} with +2V bias on those gates, while \textit{T1} and \textit{T5} are made repulsive at -3V, which is the same as the top ground plane bias. While voltage applied to \textit{T4} (V$_{T4}$) is always held fixed at +2V, we chose a \textit{T2} voltage (V$_{T2}$) to set a voltage difference, $\Delta$V = V$_{T2}$ - V$_{T4}$. For each measurement, $\Delta$V is chosen to make the potential well on the turnstile gates somewhat asymmetric. The voltage applied to \textit{T3} is gradually changed from +2V to -3V over 5 seconds, splitting the electron packet onto gates \textit{T2} and \textit{T4}. The electrons above \textit{T2} are then clocked out from the turnstile, measured in the sensing region and brought into a gate in the wide channel for storage. The remaining electrons above \textit{T4} are again distributed onto gates \textit{T2}, \textit{T3}, and \textit{T4} by bringing the bias on \textit{T2} and \textit{T3} back to +2V with \textit{T1} and \textit{T5} at -3V. \textit{T2} is set to the same V$_{T2}$, and the splitting sequence is repeated until no more electrons are found above \textit{T2} after splitting. Electrons above \textit{T4} are then brought back to the sensing region for measurement. These electrons are returned to those stored during the splitting and we perform the experiment again with the same group of electrons but a new $\Delta$V. Typical clock gate periods are 160 milliseconds. Given the rather low electron mobility at the operation temperature of 1.6K, hot electron effects are expected to be negligible at these time scales.

\begin{figure}[h]
\includegraphics[width=\textwidth]{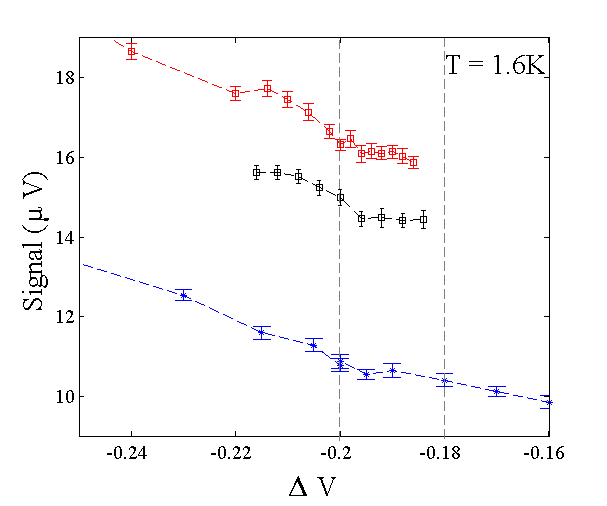}
\caption[]{\label{fig:data}Plot of electron signal left above \textit{T4} after multiple splitting of electron packets, versus $\Delta$V, the voltage difference between \textit{T2} and \textit{T4}. A plateau is seen in the region between $\Delta$V = -0.20V to -0.18V. 1V change on the \textit{T2} gate corresponds to changing the potential on the helium surface by approximately 20meV. Each dataset is from a different experiment. Dotted lines connecting data points are guide to the eye.}
\end{figure}

The signal, proportional to the number of electrons which were found above gate \textit{T4} at the end of the splitting process integrated across all seventy-eight channels, is shown in Fig.\ref{fig:data} as a function of $\Delta$V. The asymmetry while splitting the electron packets with varying $\Delta$V determines the number of electrons left in \textit{T4}. When there are many electrons in the turnstile approximately half of the electrons go onto \textit{T2} and the other half onto \textit{T4}. After multiple ``splits'' there will be two electrons left, and the Coulomb repulsion assists the splitting of the electrons onto \textit{T2} and \textit{T4}, as long as $\Delta$V is small (in magnitude) so the two electrons will prefer to split onto two gates, rather than both electrons residing above \textit{T4}. This sets the lower bound of V$_{T2}$ which would allow one electron on \textit{T4}. When there is one electron in a channel, the electron will move onto \textit{T4} as long as $\Delta$V is large enough for the last electron to fall onto \textit{T4} instead of \textit{T2}. This sets the upper bound of V$_{T2}$, which would allow the final electron to go onto \textit{T4} after splitting. A plateau is seen in Fig.\ref{fig:data} over multiple experiments between $\Delta$V = -0.20 and -0.18V. From finite element calculations, we find that applying 1V on either \textit{T2} or \textit{T4} changes the potential on the helium surface by 20meV and 1V on \textit{T3} changes the potential by 15meV. For most experiments, 4mV steps were taken on \textit{T2} gate, which corresponds to about 0.08mV (0.9K) on the surface. All the experiments are done at 1.6K, so the plateaus shown in Fig.\ref{fig:data} are a few kT wide. 

\section{Conclusion}

The existence of a plateau suggests a quantized electron signal, where the number of electrons per channel is fixed. Signal calibration using the measured capacitance (room temperature) and an estimate of the induced voltage on the \textit{sense} gate with the parameters used in the experiments gives a signal of approximately 8.5$\mu$V for one electron per channel when all seventy-eight channels are filled. The measured signal at the plateau region is between 10$\mu$V to 16$\mu$V. These values are larger than our estimate for a single electron per channel signal. If we take into account that not all the channels are filled with electrons, the measured signal approximately corresponds to two electrons per channel. The uniformity of electron occupation across the channels has been measured using a vertical channel as previously demonstrated \cite{Takita_2012}. We find that the number of channels filled with electrons varies from cooldown to cooldown with some empty channels at the top and bottom of the array. The smaller signal for the blue set of data points in Fig.\ref{fig:data}, as compared to the other two experiments probably arises from a larger number of empty channels in that experiment. From these considerations we conclude that the plateaus probably correspond to two electrons in a channel.

To investigate whether the plateaus in this region result from having one or two electrons per channel, detailed measurements at smaller $\Delta$V are necessary. Nonetheless, the plateaus seen in the measurements are a clear indication of the existence of a regime in which the number of electrons in a channel is always fixed, and that all the channels are operating simultaneously to quantize the charge with one set of gate voltages.

\section*{Acknowledgments}
	We would like to thank J. Donnal for the electronics to apply clock voltages to the CCD. Work at Princeton was supported by the NSF under Grant No. DMR-1005476.

\bibliography{arxiv_isolating}

\end{document}